\documentclass[%
 aip,
 amsmath,amssymb,
 reprint,%
]{revtex4-1}

\usepackage{graphicx}
\usepackage{dcolumn}
\usepackage{bm}
\usepackage[utf8]{inputenc}
\usepackage[T1]{fontenc}
\usepackage{mathptmx}
\usepackage{etoolbox}
\usepackage{tikz}
\usetikzlibrary{arrows.meta, positioning, shapes}

\makeatletter
\def\@email#1#2{%
 \endgroup
 \patchcmd{\titleblock@produce}
  {\frontmatter@RRAPformat}
  {\frontmatter@RRAPformat{\produce@RRAP{*#1\href{mailto:#2}{#2}}}\frontmatter@RRAPformat}
  {}{}
}%
\makeatother

\begin{document}

\title[Emergent E-I Structure in Evolved Reservoirs]{Emergent E-I 
Structure in Performance-Evolved Reservoir Networks of Neuronal 
Population Dynamics}

\author{Manish Yadav}
\email{manish.yadav@tu-berlin.de}
\affiliation{Chair of Cyber-Physical Systems in Mechanical Engineering, 
Technische Universit\"{a}t Berlin, Stra\ss e des 17. Juni 135, 
10623 Berlin, Germany}

\date{\today}

\begin{abstract}
Understanding how network structure gives rise to neuronal dynamics and whether compact computational models can recover that structure from data alone is a central challenge in computational neuroscience. We apply the performance-dependent network evolution (PDNE) framework to model the dynamics of the Wilson-Cowan (WC) neuronal system, a canonical two-population model of excitatory-inhibitory (E-I) 
interaction underlying physiological rhythms. Starting from a minimal 
seed network, PDNE iteratively grows and prunes a reservoir computing 
(RC) network based solely on prediction performance, yielding compact, 
task-optimized reservoirs networks. The evolved networks 
accurately predict both excitatory $E(t)$ and inhibitory $I(t)$ 
population activities across unseen stimulus amplitudes and 
generalize in a zero-shot manner to novel stimulus configurations:
varying pulse number, position and amplitude without retraining. Structural analysis of the evolved networks reveals a consistent 
functional organization with 
nodes specialized for E, I, and shared E-I representations. 
Importantly, the population-level connectivity of the evolved 
reservoirs spontaneously recovers the correct excitatory-inhibitory 
sign pattern of the WC model for three of four interaction types, 
without this being imposed by design. These results demonstrate that 
performance-driven network evolution can produce not only accurate 
but structurally interpretable models of physiological rhythms, 
opening a path toward compact, data-efficient digital twins of 
neuronal systems.
\end{abstract}

\maketitle

\section{Introduction}\label{sec:intro}

The brain is a paradigmatic example of a complex physiological 
network whose emergent rhythms from slow delta oscillations to fast 
gamma bursts, arising from the interplay of excitatory and inhibitory 
neuronal populations \cite{Buzsaki2004, Wang2010}. Understanding how 
network structure gives rise to these dynamics, and conversely, how 
dynamical function constrains network topology, is a central challenge 
in computational neuroscience and brain-inspired computing 
\cite{Sporns2011}. Recent advances in machine learning have opened 
new avenues for data-driven modeling of such physiological rhythms, 
yet dominant deep learning frameworks remain largely black-box 
systems: they predict well but offer little insight into the 
mechanisms underlying the dynamics they approximate \cite{Rudin2019}. 
This tension between predictive power and interpretability motivates 
the search for smaller, more structured, and mechanistically 
transparent computational models.

Neuronal population models, such as the Wilson-Cowan (WC) model 
\cite{Wilson1972}, capture the essential dynamics of 
excitatory--inhibitory (E-I) interactions with minimal 
parameterization, producing rich dynamical repertoires including 
fixed points, limit cycles, and bistability. Reproducing such 
dynamics with machine learning models has been attempted with 
recurrent neural networks and neural ordinary differential equations 
\cite{Rackauckas2020, Durstewitz2023}, but these approaches typically 
require large numbers of parameters and provide limited structural 
interpretability. A key open question is whether a compact, learnable 
dynamical system can not only predict neuronal population activity 
across varying stimulus conditions, but also develop an internal 
structure that reflects the underlying E-I organization of the target 
system.

\begin{figure*}
\includegraphics[width=0.99\textwidth]{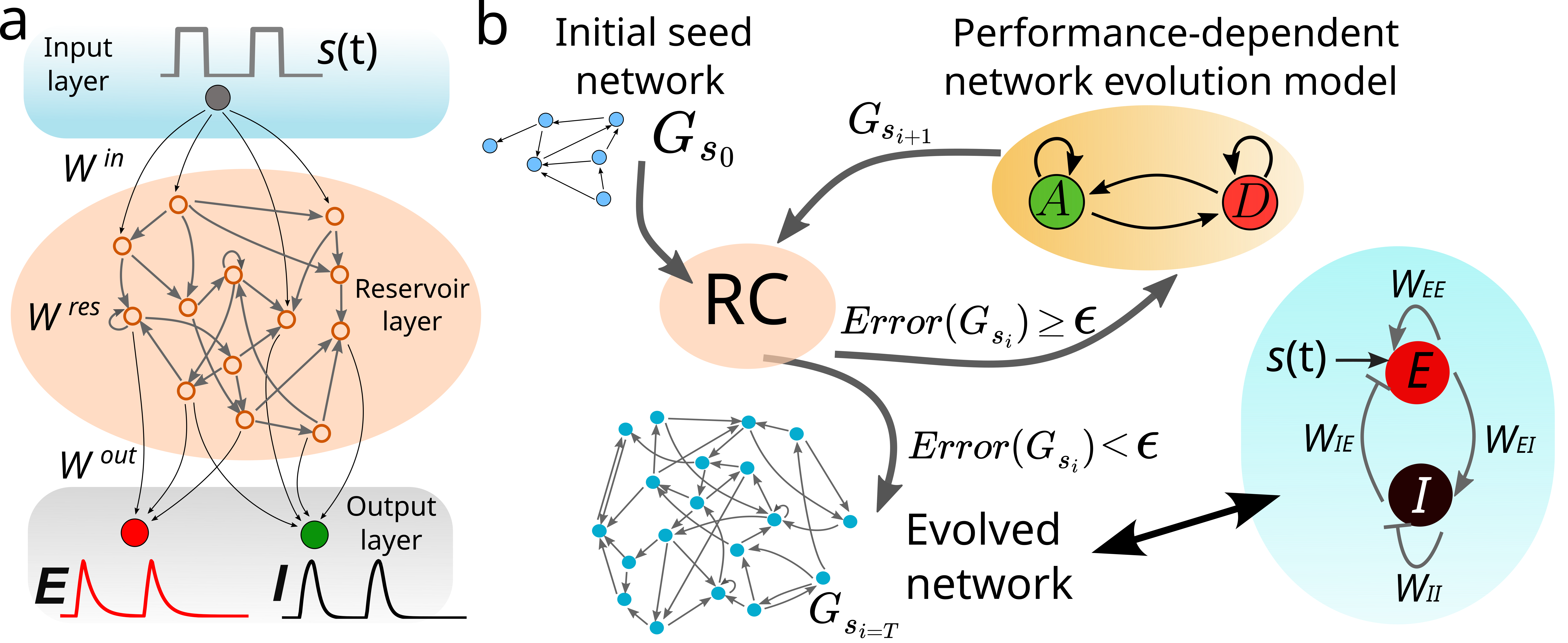}
\caption{\textbf{Performance-dependent network evolution for learning 
Wilson-Cowan neuronal dynamics.} (a)~The reservoir computing (RC) 
model with input layer receiving pulse stimulus $s(t)$, recurrent 
reservoir layer $W^{\text{res}}$, and output layer predicting 
excitatory ($E$, red) and inhibitory ($I$, green) population 
activities. (b)~The PDNE algorithm iteratively evolves the reservoir 
via node addition ($A$) and deletion ($D$) modules driven by 
prediction performance, from a minimal seed network $G_{s_0}$ toward 
a task-optimized topology. Right: the target WC model with coupling 
weights $W_{EE}$, $W_{EI}$, $W_{IE}$, and $W_{II}$.}
\label{fig1}
\end{figure*}

Reservoir computing (RC) \cite{Jaeger2001, Maass2002} offers a 
promising framework in this direction. Grounded in dynamical systems 
theory, RC exploits the rich transient dynamics of a fixed recurrent 
network, the reservoir, to project input signals into a 
high-dimensional state space, from which outputs are read out via 
simple linear regression (Fig.~\ref{fig1} (a)). This separation of 
dynamics and learning makes RC both computationally efficient and 
analytically tractable. For discrete-time systems, the reservoir 
state evolves as:
\begin{equation}
    \mathbf{r}_{t+1} = (1 - \alpha)\,\mathbf{r}_t + 
    \alpha\, G\!\left(\mathbf{A}\mathbf{r}_t + 
    \mathbf{W}_{\mathrm{in}}\mathbf{u}_t\right)
    \label{eq:rc}
\end{equation}
where $\mathbf{r}_t$ is the reservoir state, $\mathbf{u}_t$ the 
external input, $\mathbf{A}$ the recurrent adjacency matrix, $G$ a 
nonlinear activation function, and $\alpha \in (0,1]$ the leak rate. 
Only the linear readout weights $\mathbf{W}_{\mathrm{out}}$ are 
trained, making the approach highly efficient. However, the reservoir 
topology, typically random and fixed, which remains opaque, as there is no guarantee that its structure bears any relation to the dynamical 
system it is modeling. This raises a fundamental limitation: while RC 
can approximate complex dynamics, the \textit{structure-function} 
relationship of the reservoir remains uninformative by construction. 
Increasing reservoir size does not reliably improve performance 
\cite{PDNE_Yadav} and can introduce chaotic dynamics or 
signal degradation \cite{Verstraeten2007, Gallicchio2017}. Systematic 
strategies for building smaller, task-optimized reservoirs are 
therefore essential, both for computational efficiency and for the 
prospect of structural interpretability \cite{Lukosevicius2012, 
Schrauwen2009}.

A step toward this goal was taken by Yadav et al.\ \cite{PDNE_Yadav}, 
who introduced the performance-dependent network evolution (PDNE) 
framework: a biologically inspired mechanism that grows a reservoir 
from a minimal seed network by iteratively adding and pruning nodes 
based on prediction performance. Rather than fixing the topology 
before training, PDNE evolves the network structure alongside its 
function, yielding compact reservoirs adapted to the complexity of 
the target task. Related ideas have been explored in deep neural 
networks through progressive growth strategies 
\cite{DeepNN_Grow_Sudeshna} and structured pruning 
\cite{Pruning_Yadav}, demonstrating that performance-driven 
structural adaptation can improve both efficiency and generalization. 
Crucially, networks evolved under task-specific pressure may develop 
structural motifs that are not arbitrary but functionally meaningful, 
opening a path toward interpretable, structure-aware dynamical models.

In this work, we apply the PDNE framework to model the dynamics of 
the Wilson-Cowan neuronal system, a canonical model of E-I 
population dynamics underlying physiological rhythms. The WC model 
describes the mean activities of an excitatory population $E(t)$ and 
an inhibitory population $I(t)$ as:
\begin{equation}
    \tau_E \frac{dE}{dt} = -E + S_E\!\left(w_{EE}E - w_{EI}I + 
    s(t)\right)
    \label{eq:wc_E}
\end{equation}
\begin{equation}
    \tau_I \frac{dI}{dt} = -I + S_I\!\left(w_{IE}E - w_{II}I\right)
    \label{eq:wc_I}
\end{equation}
where $S_X$ is a sigmoidal activation function:
\begin{equation}
    S_X(x) = \frac{1}{1 + \exp\!\left[-a_X(x - \theta_X)\right]}, 
    \qquad X \in \{E, I\}
    \label{eq:sigmoid}
\end{equation}
and $s(t)$ is an external pulse stimulus acting as the bifurcation 
parameter. Despite its simplicity, just two coupled equations of the 
WC model exhibits a rich dynamical repertoire, including 
monostability, bistability, and sustained oscillations, making it an 
ideal testbed for structure-aware dynamical modeling. All model 
parameters are provided in Appendix~\ref{appendix_parameter}.

Our central hypothesis is two-fold. First, that PDNE-evolved 
reservoirs can accurately predict WC population dynamics across 
unseen stimulus amplitudes and stimulus configurations, generalizing 
across the system's bifurcation landscape with far fewer parameters 
than standard large reservoirs. Second, and more fundamentally, that 
the topology of the evolved network will spontaneously reflect the 
E-I connectivity structure of the WC system is not by design, but as 
an emergent consequence of performance-driven structural adaptation. 
If realized, this would demonstrate that task-optimized network 
evolution can produce not just accurate but structurally 
interpretable models of physiological rhythms, where the learned 
network architecture encodes mechanistic information about the target 
system.

\begin{figure*}
    \centering
    \includegraphics[width=\textwidth]{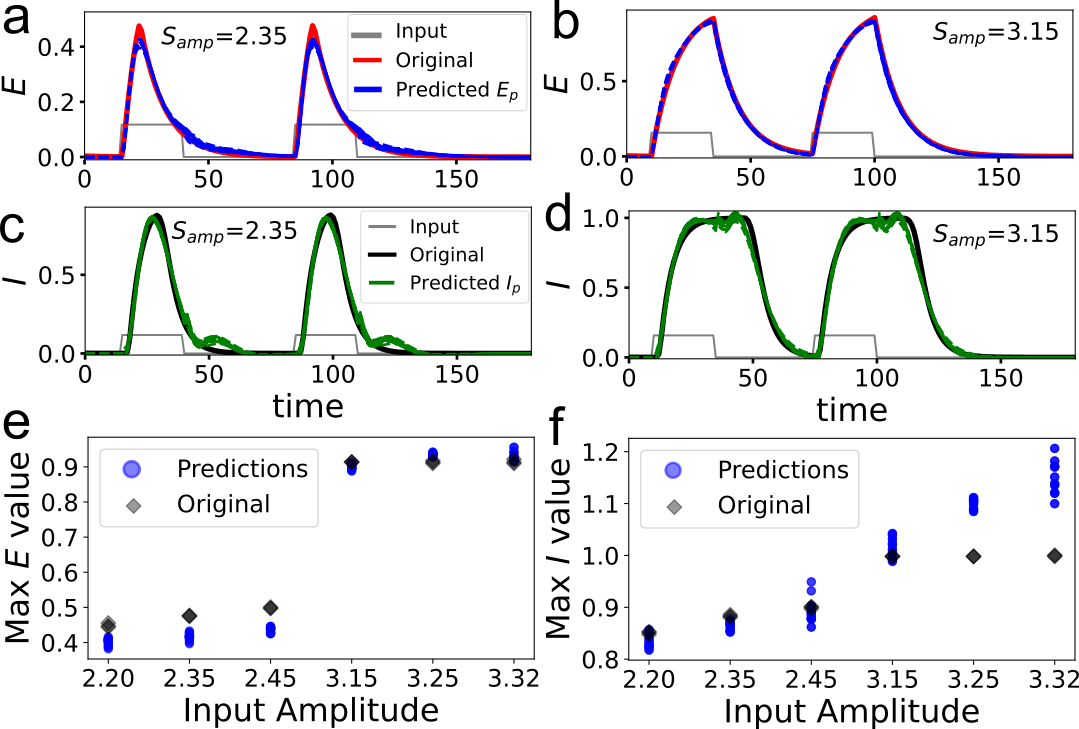}
    \caption{\textbf{Prediction performance on unseen test stimuli.} 
    \textit{Top two rows:} Predictions of excitatory $E(t)$ 
    (red: original, blue: $E_p$) and inhibitory $I(t)$ 
    (black: original, green: $I_p$) for test amplitudes 
    $S_{\mathrm{amp}} = 2.35$ (left) and $S_{\mathrm{amp}} = 3.15$ 
    (right), both absent from the training set.
    \textit{Bottom row:} Peak amplitude predictions (blue circles) 
    vs.\ original peak values (gray diamonds) across 10 model 
    repetitions. Arrows indicate systematic over- or 
    under-estimation where present. The pulse input $s(t)$ is 
    scaled by $1/20$ for visualization.}
    \label{fig2}
\end{figure*}

\section{Methods: RC Implementation and PDNE}\label{sec:methods}

\subsection{Reservoir Computing Implementation}

Building on the RC framework of Eq.~\eqref{eq:rc}, here we specify 
the implementation used for modeling WC dynamics. The activation 
function is $G = \tanh$, the recurrent matrix $W^{\text{res}} \in 
\mathbb{R}^{N \times N}$ is rescaled to spectral radius $\rho$, and 
a node-wise gain or scaling vector $\mathbf{g} \in \mathbb{R}^N$ introduces 
heterogeneous activation sensitivities across nodes. Input weights 
$W^{\text{in}} \in \mathbb{R}^{N}$ are nonzero only at designated 
input nodes $\mathcal{I} \subseteq \{1,\ldots,N\}$. The predicted 
outputs are computed as linear readouts from designated output node 
subsets $\mathcal{O}_i \subseteq \{1,\ldots,N\}$ for each channel 
$i \in \{E, I\}$:
\begin{equation}
    \hat{y}_i(t) = W^{\text{out}}_i\, \mathbf{r}_{\mathcal{O}_i}(t)
\end{equation}
Output weights are obtained via ridge regression over all training 
batches simultaneously:
\begin{equation}
    W^{\text{out}}_i = V_i R_{\mathcal{O}_i}^\top 
    \!\left(R_{\mathcal{O}_i} R_{\mathcal{O}_i}^\top + 
    \beta\, \mathbf{I}\right)^{-1}
    \label{eq:ridge}
\end{equation}
where $R_{\mathcal{O}_i} \in \mathbb{R}^{|\mathcal{O}_i| \times BT}$ 
is the matrix of reservoir states at output nodes concatenated across 
all $B$ training batches and $T$ timesteps, $V_i$ the corresponding 
target output matrix, and $\beta$ the regularization parameter.

\subsection{Performance-Dependent Network Evolution}

Standard RC fixes the reservoir topology before training, limiting 
adaptability to the complexity of the target dynamics. We adopt the 
PDNE framework \cite{PDNE_Yadav}, in which the reservoir topology is 
grown iteratively until a target prediction error is achieved. 
Starting from a minimal seed network of $N_0$ nodes, the algorithm 
alternates between two structural operations, node addition and 
deletion (Fig.~\ref{fig1}b) at each evolutionary step $t$:

\paragraph{\textbf{Node Addition ($A$).}}
A candidate node is proposed with $k \sim \mathcal{U}\{1, k_{\max}\}$ 
random directed connections to existing nodes, with weights 
$w \sim \mathcal{U}(-1, 1)$ and connection direction governed by 
probability $\Psi$. A node-wise gain $g \sim \mathcal{U}(0.01, 1.0)$ 
is assigned. The candidate is added to input set $\mathcal{I}$ with 
probability $P_{\text{inp}}$ and independently to each output set 
$\mathcal{O}_i$ with probability $P_{\text{out}}$. Addition is 
accepted if the test NMSE strictly decreases across all channels
    $\varepsilon^{\text{pred}}_{\text{new}} < 
    \varepsilon^{\text{pred}}_{\text{old}}$
otherwise a new candidate is drawn, up to $T_{\text{add}}$ attempts.

\paragraph{\textbf{Node Deletion ($D$).}}
Once any output channel satisfies NMSE $< \Delta_\varepsilon$, a pruning phase is triggered. Up to 
$\lfloor p_{\text{del}}\, N / 100 \rfloor$ deletion attempts are 
made, each randomly selecting a node for removal. Node indices are 
remapped after each deletion to maintain contiguous indexing. 
Deletion is accepted if the test NMSE does not increase on any 
channel $\varepsilon^{\text{pred}}_{\text{new}} \leq 
    \varepsilon^{\text{pred}}_{\text{old}}$
yielding a compact network retaining only task-relevant nodes. 
Evolution terminates when all channels satisfy 
$\varepsilon^{\text{pred}} \leq \Delta_\varepsilon$ or the maximum 
evolution steps $T$ is reached. The NMSE for output channel $i$ is:
\begin{equation}
    \varepsilon_i = \frac{
    \sum_t \left(\hat{y}_i(t) - y_i(t)\right)^2}{
    \sum_t \left(y_i(t) - \bar{y}_i\right)^2}
\end{equation}
Performance is evaluated as the mean NMSE across $B$ training 
amplitudes and $N_{\text{RC}}$ independent RC initializations per 
evolution step to reduce stochastic variability.

\begin{figure*}
    \centering
    \includegraphics[width=\textwidth]{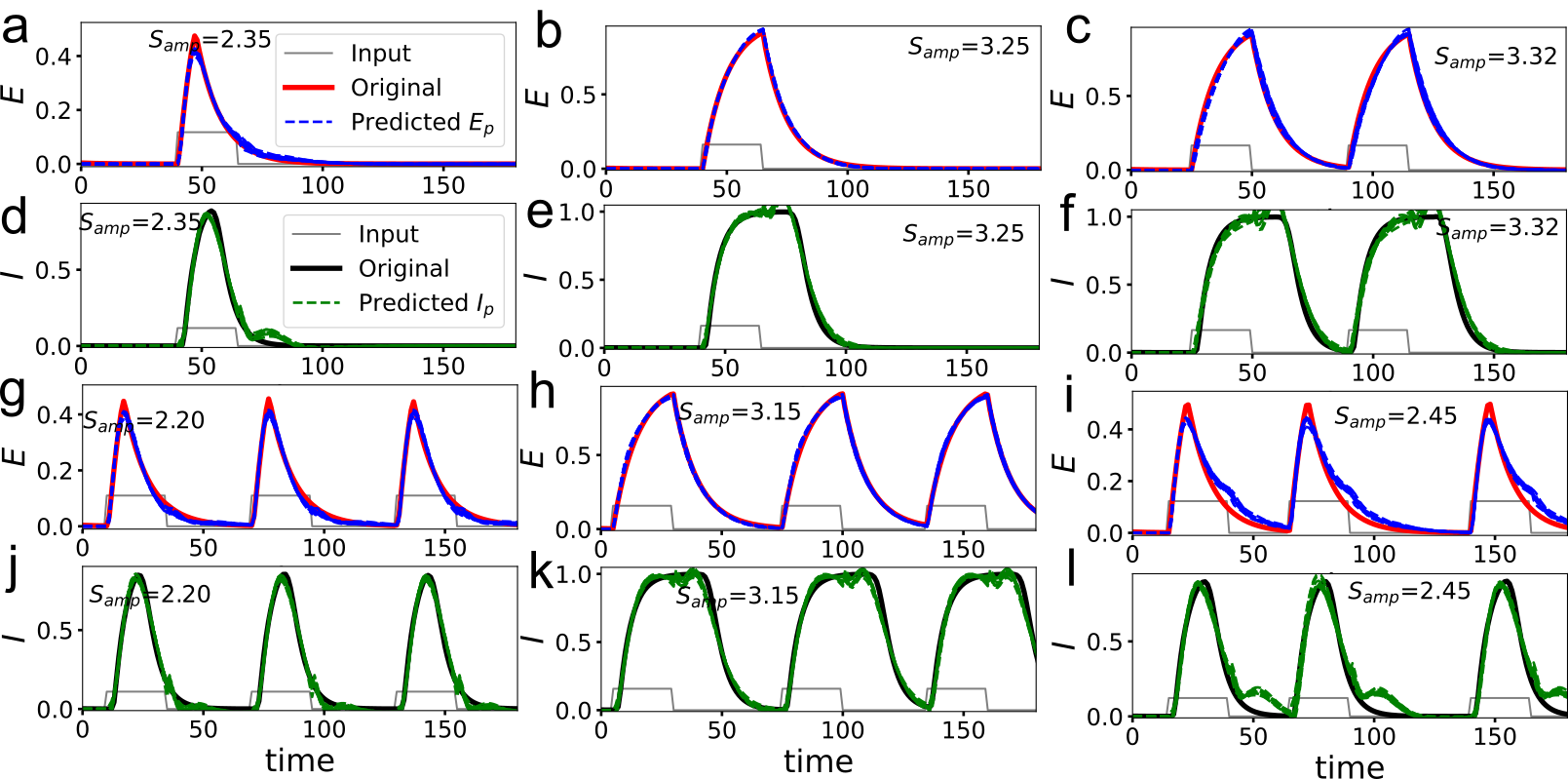}
    \caption{\textbf{Zero-shot generalization to novel stimulus 
    configurations.} Models trained exclusively on two-pulse 
    sequences are tested on stimuli with varying number of pulses, 
    positions, and amplitudes without retraining. Each panel pair 
    shows $E(t)$ (top: red original, blue dashed $E_p$) and $I(t)$ 
    (bottom: black original, green dashed $I_p$).
    \textit{(a,b)} Single-pulse reference at 
    $S_{\mathrm{amp}} = 2.35$ and $3.25$.
    \textit{(c--l)} Multi-pulse stimuli at 
    $S_{\mathrm{amp}} = 2.20,\, 2.45,\, 3.15,\, 3.25,\, 3.32$ 
    with varying pulse counts and temporal positions. The pulse 
    input $s(t)$ is scaled by $1/20$ for visualization.}
    \label{fig3}
\end{figure*}

\subsection{Training Data}

We simulate the WC system under pulse stimuli of fixed duration 
$T_{\text{pulse}}$ and varying amplitude. Training amplitudes 
$\mathcal{A}_{\text{train}} = \{1.25, 1.5, 2.0, 2.5, 3.0\}$ span 
the excitable and oscillatory regimes, while test amplitudes 
$\mathcal{A}_{\text{test}} = \{0.85, 1.4, 1.75, 2.25, 2.75\}$ are 
interleaved between training values to probe interpolation 
generalization. Each simulation is subsampled by a factor of 10 to 
obtain RC input-output sequences. All hyperparameters are summarized 
in Appendix~\ref{appendix_parameter}.

\section{Prediction Performance and Generalization}\label{sec:results}

\subsection{Generalization Across Stimulus Amplitudes}

A key requirement for a faithful model of the WC system is the 
ability to generalize beyond the training distribution, predicting 
population dynamics for stimulus amplitudes not seen during 
evolution. Fig.~\ref{fig2} shows predictions of the final evolved 
reservoir on two test amplitudes, $S_{\mathrm{amp}} = 2.35$ and 
$S_{\mathrm{amp}} = 3.15$, both interleaved between training values 
and withheld throughout the PDNE procedure. In this open-loop 
setting, only the external pulse stimulus $s(t)$ drives the 
reservoir at each timestep; predicted outputs $\hat{E}(t)$ and 
$\hat{I}(t)$ are computed purely from the evolved reservoir state 
without any feedback, making accurate prediction contingent entirely 
on the internalized dynamics of the evolved topology.

The top rows of Fig.~\ref{fig2} show that the evolved network 
accurately reproduces both population responses, capturing the 
transient peak, characteristic decay timescale, and return to 
baseline under both stimulus conditions. The inhibitory predictions 
closely track $I(t)$ across both amplitudes, including the faster 
rise and sharper peak relative to $E(t)$, a direct consequence of 
the shorter inhibitory timescale $\tau_I < \tau_E$ in the WC model. 
The excitatory predictions show slightly larger deviations near the 
peak at $S_{\mathrm{amp}} = 3.15$, consistent with increased 
response nonlinearity at higher stimulus amplitudes.

The bottom row provides a quantitative assessment of peak amplitude 
prediction across 10 independent model repetitions. For 
$S_{\mathrm{amp}} = 2.35$, predictions cluster tightly around the 
original peak values, indicating consistent generalization. For 
$S_{\mathrm{amp}} = 3.15$, greater spread is observed, reflecting 
the increased variability of stochastic network evolution at higher 
response amplitudes. Notably, the reservoir reproduces $I(t)$ 
accurately without ever receiving $I$ as an input, relying solely on $s(t)$, suggesting that E-I coupling dynamics have been 
implicitly encoded in the evolved topology, a point examined 
structurally in Section~\ref{sec:structure}.

\begin{figure*}[t]
    \centering
    \includegraphics[width=0.99\textwidth]{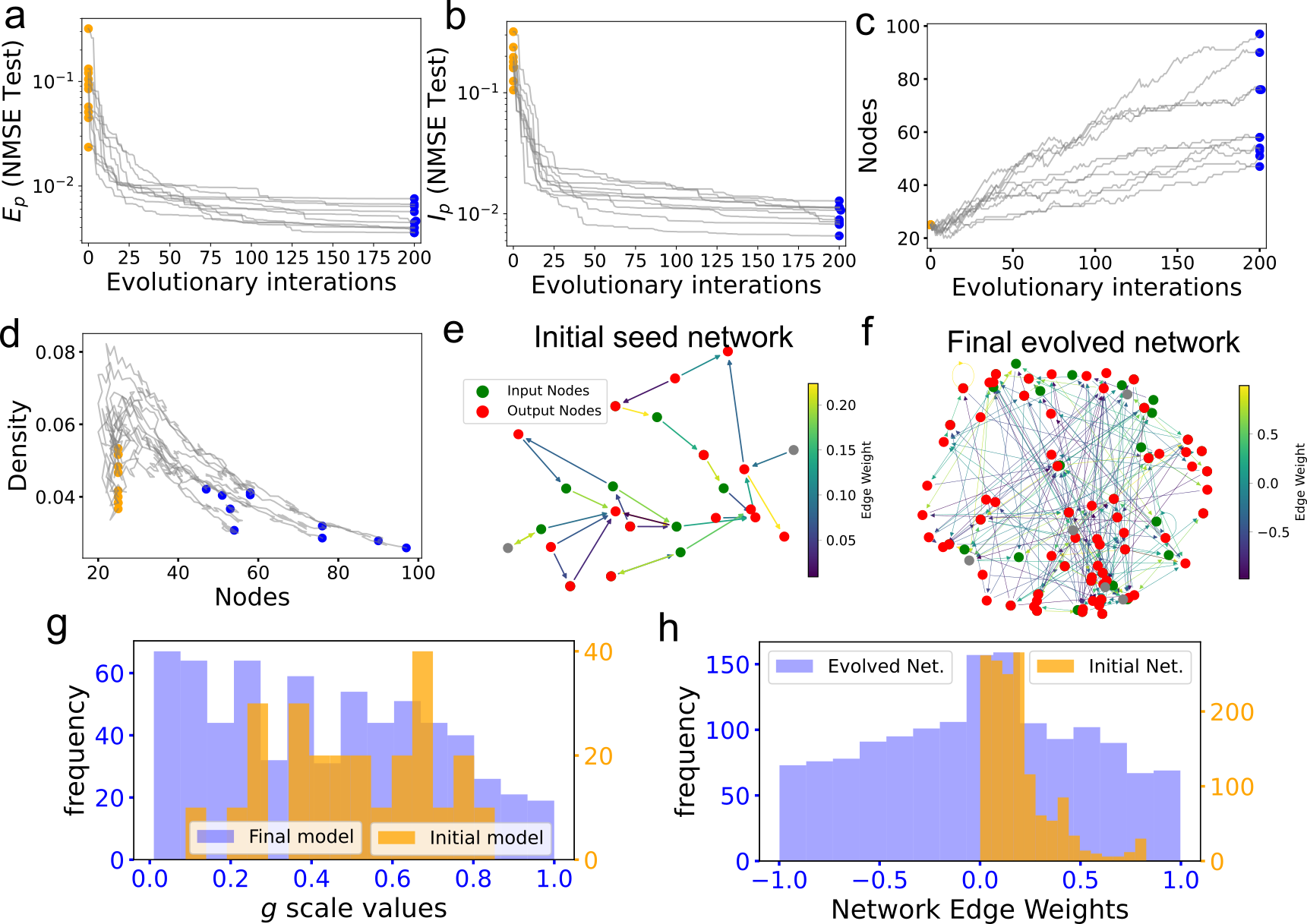}
    \caption{\textbf{PDNE evolution dynamics and network topology 
    across 10 model repetitions.} 
    (a,b)~Training NMSE for $\varepsilon_E$ and $\varepsilon_I$ 
    vs.\ evolution step $t$. Orange: initial seed; blue: final 
    evolved network.
    (c)~Reservoir node count $N(t)$ reflecting iterative addition 
    and deletion.
    (d)~Evolution trajectory in the density-node parametric space.
    (e,f)~Representative initial and final network topologies. 
    Green: input nodes $\mathcal{I}$; red: output nodes 
    $\mathcal{O}_{E,I}$; gray: hidden nodes. Edge colors: 
    connection weight (colorbar).
    (g,h)~Distributions of node gain values $\mathbf{g}$ and 
    edge weights for initial (orange) and evolved (blue) networks.}
    \label{fig4}
\end{figure*}

\subsection{Zero-Shot Generalization to Novel Stimulus Configurations}

A more stringent test of whether a learned model has internalized 
the underlying system dynamics rather than memorizing 
stimulus-response mappings, is its ability to generalize to 
structurally different inputs without retraining. 
Fig.~\ref{fig3} presents such a test: the PDNE-evolved reservoirs, 
trained exclusively on two-pulse sequences, are evaluated on 
entirely novel configurations varying in number of pulses, temporal 
positions, and amplitudes, with no retraining or fine-tuning.

Fig.\ref{fig3}(a,b) panels show predictions for a single-pulse stimulus at 
$S_{\mathrm{amp}} = 2.35$ and $3.25$ configurations structurally 
simpler than the training sequences confirming that the model does 
not require multiple pulses to produce accurate responses. Panels 
(c-l) demonstrate predictions across two- and three-pulse 
configurations spanning a range of amplitudes and temporal 
arrangements. The evolved reservoir accurately tracks both $E(t)$ 
and $I(t)$ across all configurations, correctly reproducing the 
response to each individual pulse, the amplitude scaling with 
stimulus strength, and the recovery dynamics between successive 
pulses.\\

This result carries an important implication: the reservoir has not learned a fixed input-output map tied to the specific two-pulse 
training format, but has instead internalized the reactive dynamics 
of the WC system, its ability to respond to an arbitrary pulse at 
any point in time with the correct excitatory and inhibitory 
trajectory. Each pulse effectively resets the system into a 
transient response, and the evolved network reproduces this 
reset-and-respond behavior across varying inter-pulse intervals and amplitudes. These findings demonstrate that PDNE-evolved reservoirs 
develop a generalizable internal model of WC dynamics, rather than a stimulus-specific approximation.

\section{Network Evolution and Structural Analysis}
\label{sec:structure}

We next analyze how the network structure evolves while solving the 
WC modeling task, tracking prediction error, network size, 
connection density, and internal parameter distributions jointly as 
the reservoir grows and is pruned.

\begin{figure*}[t]
    \centering
    \includegraphics[width=\textwidth]{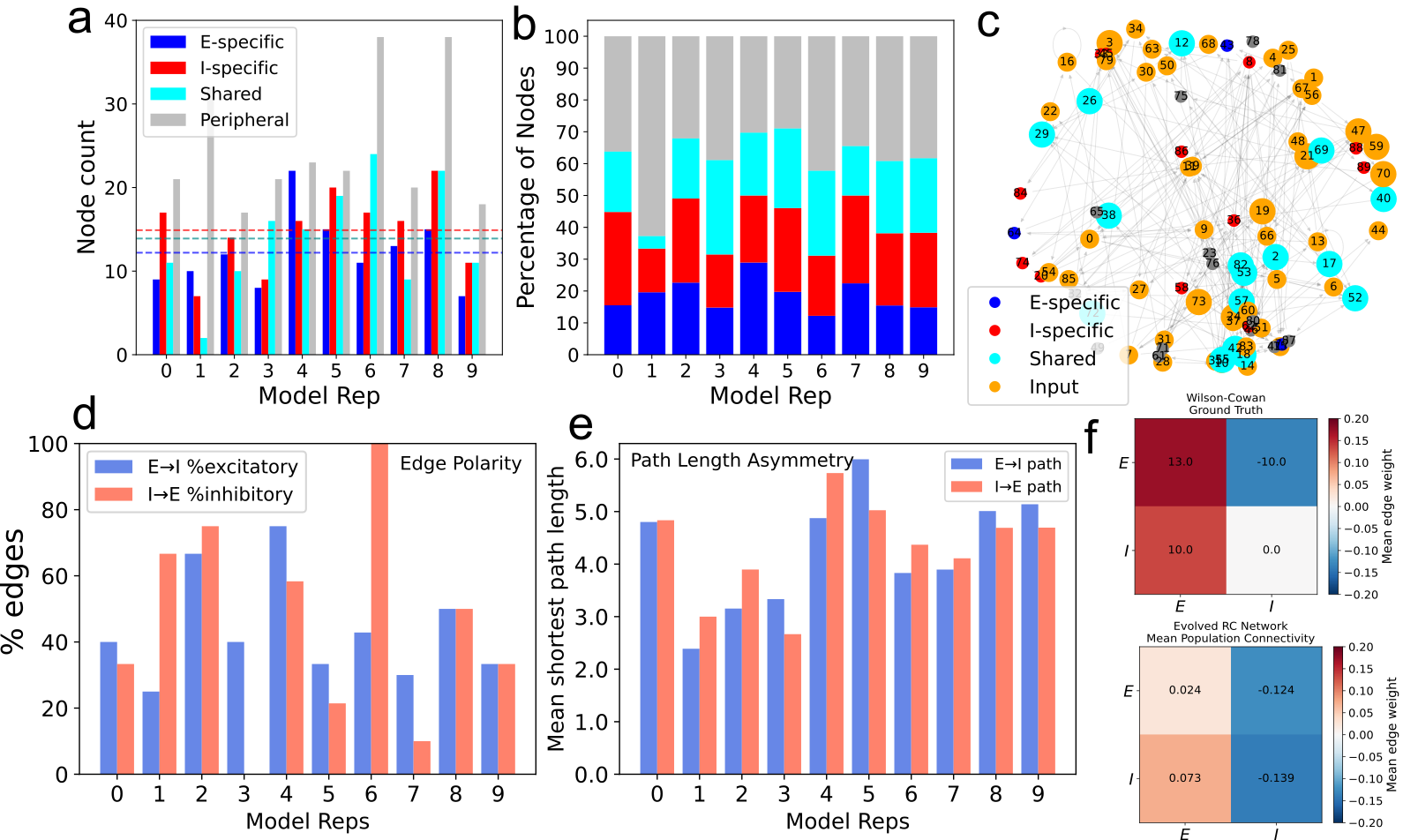}
    \caption{\textbf{Structural analysis of evolved reservoirs and 
    comparison with the Wilson-Cowan E-I architecture.}
    (a)~Node classification counts per repetition: E-specific 
    (blue), I-specific (red), Shared (cyan), Peripheral (gray). 
    Dashed lines: cross-repetition means.
    (b)~Stacked percentage node composition per repetition.
    (c)~Representative evolved network colored by functional role: 
    E-specific (blue), I-specific (red), Shared (cyan), Input 
    (orange).
    (d)~Edge polarity: \% of E$\rightarrow$I connections that are 
    excitatory (blue) and I$\rightarrow$E connections that are 
    inhibitory (red) per repetition.
    (e)~Mean shortest path lengths E$\rightarrow$I (blue) and 
    I$\rightarrow$E (red) per repetition.
    (f)~WC ground truth connectivity matrix (top) and mean evolved 
    network population connectivity matrix (bottom), showing sign 
    correspondence for three of four E-I interaction types.}
    \label{fig5}
\end{figure*}

\subsection{Evolution Dynamics}

Fig.~\ref{fig4} characterizes these dynamics across 10 independent 
model repetitions. Panels (a) and (b) show the training NMSE for 
the excitatory ($\varepsilon_E$) and inhibitory ($\varepsilon_I$) 
channels as a function of evolution step $t$. All repetitions begin 
from the same minimal seed (orange markers) and converge toward 
$\Delta_\varepsilon$ (blue markers) via qualitatively consistent but 
stochastically distinct trajectories. The rapid initial error 
reduction reflects the high marginal utility of early node 
additions, while convergence slows as the network approaches the 
target threshold.

The number of reservoir nodes $N(t)$ are shown in fig.\ref{fig4} (c), whose 
sawtooth-like trajectory reflects the interplay of addition and 
deletion: $N(t)$ grows as new nodes reduce prediction error, then 
decreases as redundant nodes are pruned. Final network sizes vary 
across repetitions ($\bar{N} = 66 \pm 17$ nodes), reflecting the 
existence of multiple compact configurations capable of achieving 
the target performance. Panel (d) shows the joint evolution trajectory in the 
density-node parametric space. Starting from a dense but minimal seed, the 
network expands while density decreases as nodes are added faster 
than connections, before settling into a consistent region across 
repetitions. This convergence suggests the PDNE algorithm discovers 
a task-specific structural regime rather than growing the network 
arbitrarily.

The representative initial and final network 
topologies are shown in Fig.\ref{fig4} (e, f). The sparse seed network develops into a substantially 
richer recurrent structure with broader connectivity and a greater 
proportion of output nodes. Edge colors reveal the spontaneous 
emergence of both excitatory and inhibitory connections in the 
evolved network. This structural reorganization is quantified in 
panels (g) and (h). The node gain distribution broadens 
substantially after evolution from a narrow initial range to 
nearly the full admissible interval, introducing heterogeneous 
activation sensitivities that enrich reservoir dynamics. 
Correspondingly, the edge weight distribution restructures from 
one dominated by weak positive connections to a broad, 
near-symmetric distribution spanning both positive and negative 
values. The increased prevalence of negative weights corresponding 
to inhibitory interactions, is consistent with the E-I coupling 
structure of the target WC system, suggesting that 
performance-driven evolution sculpts the reservoir topology toward 
dynamical motifs reflecting the functional organization of the 
underlying neuronal model.

\subsection{Functional Node Organization}

We next carried out the structural analysis of the evolved 
reservoirs, examining whether the network topology reflects the E-I 
organization of the target WC system. We classify each reservoir 
node into four functional categories based on its readout weight 
contributions: E-specific (contributing primarily to $\hat{E}$), 
I-specific (contributing primarily to $\hat{I}$), Shared 
(contributing significantly to both), and Peripheral (contributing 
weakly to neither).

As shown in Fig.~\ref{fig5} (a, b), the node classification counts and percentage 
composition across 10 repetitions. Despite substantial variation in 
final network size ($\bar{N} = 66 \pm 17$), the functional 
composition remains remarkably consistent: E-specific, I-specific, 
and Shared nodes account for approximately $19\%$, $23\%$, and 
$20\%$ of the network respectively, with means of 
$\bar{n}_E = 12.2 \pm 4.2$, $\bar{n}_I = 14.9 \pm 4.5$, and 
$\bar{n}_S = 13.9 \pm 6.3$ nodes. The slight dominance of 
I-specific nodes is consistent with the faster inhibitory timescale 
$\tau_I < \tau_E$ in the WC model, which places greater 
representational demands on the inhibitory population. The large 
proportion of Shared nodes are comparable in count to either 
specialized population, reflects the tight mutual E-I coupling in 
the WC system, requiring nodes that integrate information from both populations. A representative evolved network topology colored by functional role is provided in Fig. \ref{fig5}(c). E-specific and I-specific nodes are spatially 
interleaved rather than segregated into distinct clusters, 
consistent with the distributed solution discovered by the PDNE 
algorithm.

\subsection{Structural Correspondence with Wilson-Cowan 
Connectivity}\label{sec:lateral}

Furthermore, we examine whether the evolved 
reservoir recapitulates the signed connectivity structure of the WC 
model at the population level. The edge polarity 
across repetitions are shown in Fig.~\ref{fig5} (d), where the percentage of excitatory 
E$\rightarrow$I and inhibitory I$\rightarrow$E connections varies 
and falls below the WC prediction of 100\%, the correct sign 
tendency is present in the majority of repetitions. Panel (e) shows 
mean shortest path lengths in the E$\rightarrow$I and 
I$\rightarrow$E directions, which remain largely symmetric across 
repetitions, indicating that the network does not replicate the 
fast-excitation, delayed-inhibition temporal asymmetry of the WC 
model at the topological level.

The most direct structural comparison is provided by panel (f). The 
WC ground truth connectivity matrix (top) has the sign pattern: 
$w_{EE} > 0$, $w_{EI} < 0$, $w_{IE} > 0$, $w_{II} = 0$. The 
evolved network's mean population connectivity matrix (bottom) 
recovers the correct sign for three of four interactions: 
E$\rightarrow$E ($+0.024$), E$\rightarrow$I ($-0.124$), and 
I$\rightarrow$E ($+0.073$) all match the WC signs. The sole 
deviation is I$\rightarrow$I ($-0.139$), which deviates from the WC 
value of zero through stronger lateral inhibition within the 
I-population. Importantly, this lateral I$\rightarrow$I inhibition 
($|\bar{w}| = 0.47 \pm 0.12$) is consistently stronger than 
I$\rightarrow$E inhibition ($|\bar{w}| = 0.31 \pm 0.28$) across 
$9/10$ repetitions, constituting a novel emergent inhibitory motif 
absent in the original WC formulation but consistent with known 
interneuron physiology in cortical circuits \cite{Tremblay2016}.

These results demonstrate that the PDNE evolution process, driven 
purely by prediction performance, spontaneously recovers the 
qualitative E-I connectivity structure of the Wilson-Cowan model. 
The evolved reservoir does not replicate the explicit two-node WC 
architecture, but instead develops a distributed multi-node 
solution whose population-level connectivity signs are consistent 
with the underlying neuronal coupling , providing a structurally 
interpretable, compact model of E-I physiological rhythms.

\section{Discussion}\label{sec:discussion}

This study demonstrates that performance-driven network evolution 
can produce compact, accurate, and structurally interpretable models 
of neuronal population dynamics. By applying the PDNE framework to 
the Wilson-Cowan system, we show that reservoirs evolved purely on 
prediction performance develop internal organizations that 
spontaneously reflect the E-I architecture of the target neuronal 
model, without this correspondence being imposed by design.

The prediction results establish that PDNE-evolved reservoirs of 
$66 \pm 17$ nodes generalize reliably across unseen stimulus 
amplitudes and, more strikingly, to entirely novel stimulus 
configurations varying in pulse number, position, and amplitude 
without any retraining. This zero-shot generalization indicates 
that the evolved networks have internalized the reactive dynamics of 
the WC system, its stimulus-driven transient responses, timescale 
separation between E and I populations, and recovery between 
successive pulses rather than memorizing specific training 
patterns. This has direct implications for data-efficient digital 
twinning of neuronal systems, where training data may be limited 
to a narrow range of experimental conditions.

The structural analysis reveals two complementary findings. First, 
the functional node composition approximately equal proportions of 
E-specific, I-specific, and shared nodes, consistent across 
stochastically distinct repetitions, which demonstrates that PDNE evolution reliably organizes the reservoir into populations 
mirroring the E-I structure of the target system. The consistency 
of this organization across networks of varying size suggests it reflects a fundamental property of the task rather than a 
particular network instantiation. Second, the population-level 
connectivity of the evolved networks spontaneously recovers the 
correct excitatory--inhibitory sign pattern of the WC model for 
three of four interaction types. The sole structural deviation,
stronger lateral I$\rightarrow$I inhibition relative to 
I$\rightarrow$E inhibition constitutes a novel emergent motif 
that is absent in the original WC formulation but is well-known 
in cortical interneuron physiology \cite{Tremblay2016}. This 
suggests that the evolution process discovers a broader class of 
E-I stabilization mechanisms beyond those encoded in the target 
model, potentially revealing alternative dynamical solutions that are biologically plausible.\\

These findings have several broader implications. For neuronal 
modeling, they demonstrate that structure-function relationships in 
E-I networks can be recovered from dynamics alone, without 
anatomical constraints or explicit supervision of connectivity. For 
reservoir computing, they establish that PDNE provides a principled 
path from uninformative random reservoirs to task-adapted, 
structurally interpretable networks. For RC-based digital twin construction\cite{MultiParam_NLD2024}, 
they suggest that compact evolved reservoirs can serve as 
mechanistically transparent surrogates for neuronal systems, 
capturing not only the output dynamics but also the organizational 
principles of the underlying circuit.

Several directions merit future investigation. Extending the 
framework to more complex neuronal models: multi-population 
networks, spatially structured cortical circuits, or models 
exhibiting sustained oscillations and chaos will test the 
generality of the structural correspondence finding. Incorporating biophysical constraints on network connectivity during evolution 
(e.g., Dale's principle separating excitatory and inhibitory 
neurons) may further sharpen the structural interpretability of 
the evolved reservoirs. Finally, applying PDNE to experimental 
neural time-series data, where the ground truth connectivity is 
unknown, represents the ultimate test of whether performance-driven 
evolution can serve as a tool for circuit inference from dynamics.

By bridging insights from network science, reservoir computing, and 
computational neuroscience, this work opens new avenues for 
building efficient, interpretable, and structurally meaningful 
models of physiological rhythms, a step toward compact digital 
twins of neuronal systems grounded in dynamical first principles.


\section*{Data Availability Statement}

All Python codes for numerical simulations and data generated in 
this study are publicly available at: github.com/maneesh51/PDNE-WilsonCowanNeuron \url{https://github.com/maneesh51/PDNE-WilsonCowanNeuron}

\appendix

\section{Model Parameters}\label{appendix_parameter}

\subsection*{Wilson--Cowan Neuron Model}

Parameters for the WC model 
(Eqs.~\eqref{eq:wc_E}--\eqref{eq:sigmoid}): $\tau_E = 10$, 
$\tau_I = 5$, $w_{EE} = 13$, $w_{EI} = 10$, $w_{IE} = 10$, 
$w_{II} = 0$, $a_E = 6$, $a_I = 4$, $\theta_E = 2.5$, 
$\theta_I = 2.0$. The external pulse stimulus is:
\begin{equation}
s(t)=\begin{cases}
S_{\mathrm{amp}}, & 20 < t < 80 \\
0, & \text{otherwise}
\end{cases}
\end{equation}
with training amplitudes $\mathcal{A}_{\mathrm{train}} = 
\{1.25, 1.5, 2.0, 2.5, 3.0\}$ and test amplitudes 
$\mathcal{A}_{\mathrm{test}} = \{0.85, 1.4, 1.75, 2.25, 2.75\}$.
The system is integrated using \texttt{solve\_ivp} with a time 
span of $T_{\max} = 175\,\text{s}$, integration step 
$\delta t = 0.1\,\text{s}$, and subsampled by a factor of 10 
to obtain the RC input-output sequences of length 
$N_{\mathrm{pts}} = 3000$ per trial.

\subsection*{Initial seed Network}

The initial seed network is a random directed Erd\H{o}s--R\'{e}nyi 
network of $N_0 = 25$ nodes with mean degree $K_s = 1$ and initial 
spectral radius $\rho_0 = 0.2$. Input and output node sets 
$\mathcal{I}$ and $\mathcal{O}_{E,I}$ are initialized by randomly 
selecting $\lfloor P_{\mathrm{inp}} \cdot N_0 \rfloor$ and 
$\lfloor P_{\mathrm{out}} \cdot N_0 \rfloor$ nodes respectively. 
Node gain values are drawn uniformly: 
$g_i \sim \mathcal{U}(0.01, 1.0)$.

\subsection*{PDNE and RC Hyperparameters}

All hyperparameters are fixed across all 10 model repetitions 
and summarized in Table~\ref{tab:params}.

\begin{table}[h]
\centering
\caption{PDNE and RC hyperparameters.}
\label{tab:params}
\begin{tabular}{lll}
\hline\hline
Parameter & Symbol & Value \\
\hline
\multicolumn{3}{l}{\textit{Reservoir Computing}} \\
Leak rate               & $\alpha$                  & $0.2$ \\
Spectral radius         & $\rho$                    & $0.2$ \\
Ridge parameter         & $\beta$                   & $5\times10^{-10}$ \\
Transient steps         & $T_{\mathrm{trans}}$      & $10$ \\
RC repetitions          & $N_{\mathrm{RC}}$         & $1$ \\
Number of inputs        & $N_I$                     & $1$ \\
Number of outputs       & $N_O$                     & $2$ \\[4pt]
\multicolumn{3}{l}{\textit{Network Evolution}} \\
Max evolution steps     & $T$                       & $200$ \\
Target NMSE             & $\Delta_\varepsilon$      & $0.005$ \\
Error precision         & --                        & $6$ digits \\
Max addition attempts   & $T_{\mathrm{add}}$        & $25$ \\
Max new links           & $k_{\max}$                & $5$ \\
Connection probability  & $\Psi$                    & $0.5$ \\
Input node probability  & $P_{\mathrm{inp}}$        & $0.5$ \\
Output node probability & $P_{\mathrm{out}}$        & $0.5$ \\
Node gain range         & $g \sim \mathcal{U}$      & $(0.01,\;1.0)$ \\
Deletion percentage     & $p_{\mathrm{del}}$        & $20\%$ \\
Input node type         & --                        & Non-exclusive ($0$) \\
Output node type        & --                        & Non-exclusive ($0$) \\[4pt]
\multicolumn{3}{l}{\textit{Seed Network}} \\
Number of nodes         & $N_0$                     & $25$ \\
Mean degree             & $K_s$                     & $1$ \\
Initial spectral radius & $\rho_0$                  & $0.2$ \\
\hline\hline
\end{tabular}
\end{table}

\nocite{*}
%

\end{document}